\input{aipcheck}

\documentclass[
    ,final            
  ]
  {aipproc}

\layoutstyle{6x9}

\begin{document}

\title{The Acceleration History of the Universe and the Properties of the Dark Energy}

\classification{\texttt{98.80-k}}

\author{Ruth A. Daly}{
  address={Department of Physics, Penn State University, Berks Campus, P. O. 
Box 7009, Reading, PA 19610}
}

\author{S. G. Djorgovski}{
  address={Division of Physics, Mathematics, and Astronomy, 
Caltech, MS 105-24, 
Pasadena, CA 91125}
}

\begin{abstract}
The model-independent method of using type Ia supernovae
proposed and developed by Daly \& Djorgovski (2003, 2004) has
been applied to the Riess et al. (2007) supernovae sample. 
Assuming only a Robertson-Walker metric, 
we find $q_0 = -0.5 \pm 0.13$, indicating that
the universe is accelerating today.  This result is 
purely kinematic, is independent
of the contents of the universe, and does
not require that a theory of gravity be specified. Our
model-independent method allows a determination of
$q(z)$ for a particular value of space curvature. 
When $q(z)$ transitions from negative to 
positive values, the universe transitions from an
accelerating to a decelerating state.  For zero
space curvature, we find that the 
universe transitions from acceleration to deceleration
at a zedshift of about $z_T = 0.35 {}^{+0.15}_{-0.7}$
for the Riess et al. (2007) sample.    

If a theory of gravity is specified, the supernovae
data can be used to determine the pressure, energy density,
and equation of state of the dark energy, and the potential
and kinetic energy density of a dark energy scalar field
as functions of redshift. The relevant equations from 
General Relativity are applied, and
these functions are obtained.  The results are consistent with 
predictions in the standard Lambda Cold Dark Matter model
at about the two sigma level.  

\end{abstract}

\maketitle


\section{Introduction}
Determination of the expansion history of the universe, from which we can
constrain the physical nature of its matter and energy constituents, is
the central problem of traditional cosmology. One way it can 
be addressed is through the use of a set of
coordinate distances and redshifts for some standard set of objects.  
Type Ia supernovae provide a modified standard candle 
(e.g. \cite{P93,H95}) 
that allow the
distance modulus, luminosity distance, and 
coordinate distance to each source to be determined.

In a novel, model-independent approach to this problem, it
was shown by \cite{DD03} that the first and
second derivatives of the coordinate distance with respect
to redshift could be obtained from the coordinate distances
and combined to solve for the expansion rate $H(z)/H_0$ and
acceleration rate $q(z)$ of the universe.   
The only assumptions are that the universe is described by 
a Robertson-Walker metric and has zero space curvature. 
The results are independent of the contents of the 
universe and the properties of these 
contents, and independent of whether General Relativity 
provides an accurate description of the universe. 
The assumption of zero space curvature is dropped by 
\cite{D07B}, who show that the deceleration parameter
at a redshift of zero, $q_0$, remains the same, independent
of whether space curvature is zero or non-zero. 

\begin{figure}
  \includegraphics[height=0.75\textheight]{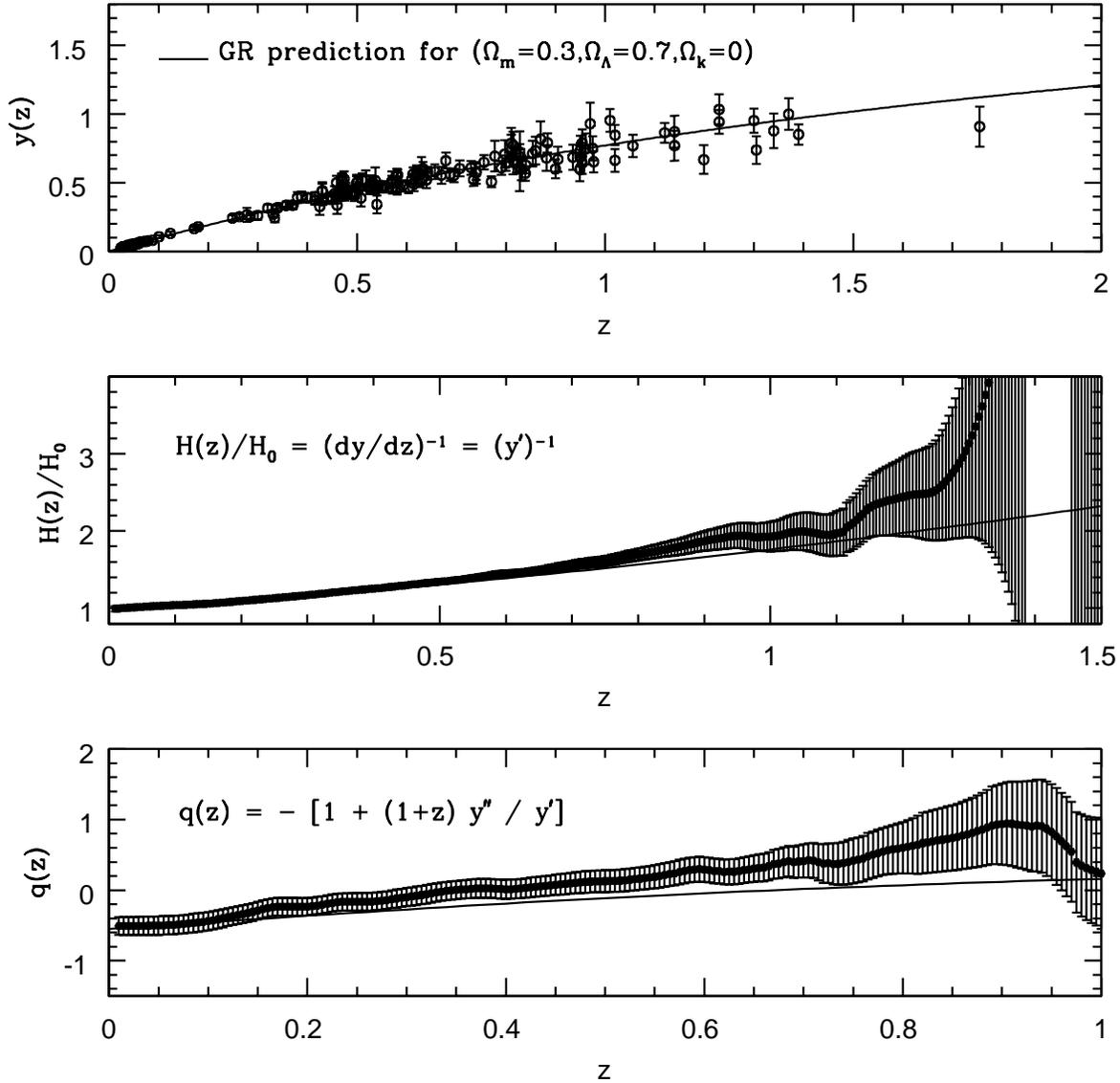}
\label{yeq}
  \caption{Dimensionless coordinate distances to 182 type Ia
supernovae of \cite{R07}, obtained with the 
best fit parameters presented by \cite{D07A}, 
are shown in the top panel.  A window function of width 0.6 in redshift
was used to obtain the first and second derivatives of the coordinate
distance with respect to redshift, as described by \cite{DD03, DD04}, and these
are used to determine the expansion and acceleration rates of the
universe as functions of redshift, shown in the second and third panels.
}
\end{figure}

\begin{figure}
  \includegraphics[height=0.75\textheight]{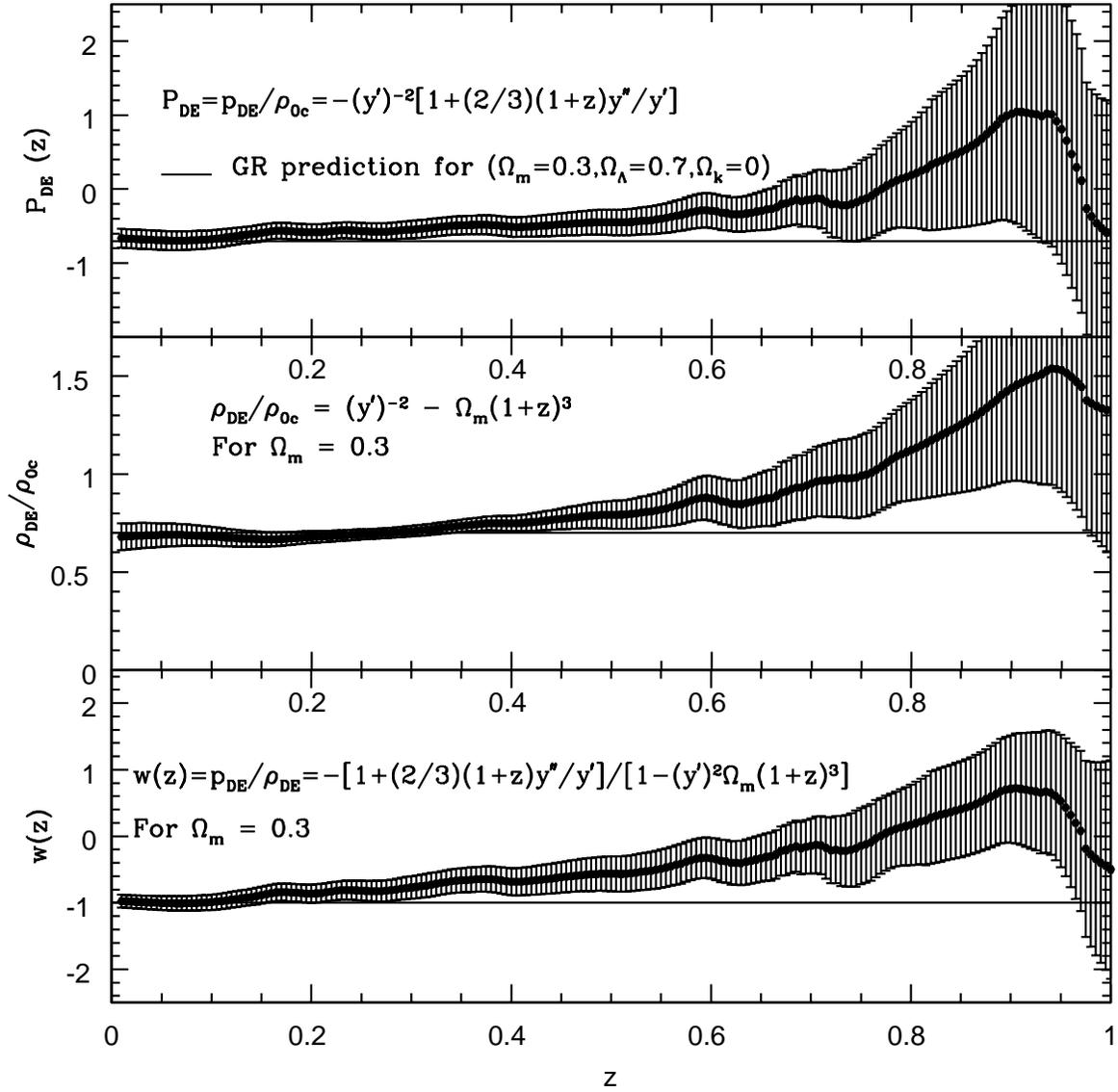}
\label{Pfw}
  \caption{The pressure, energy density, and equation of state
of the dark energy obtained by combining the first and second
derivatives of the coordinate distance for the sample of 
\cite{R07}.  As described by \cite{DD04}, a theory of gravity must
be specified to 
obtain the pressure of the dark energy as a function
of redshift, and General Relativity has been applied here. 
To obtain the energy density and equation of state of the 
dark energy as functions of redshift, the
present value of the mean mass density in non-relativistic 
matter must be specified, and we have adopted 
$\Omega_m = 0.3$.}
\end{figure}

\begin{figure}
  \includegraphics[height=0.75\textheight]{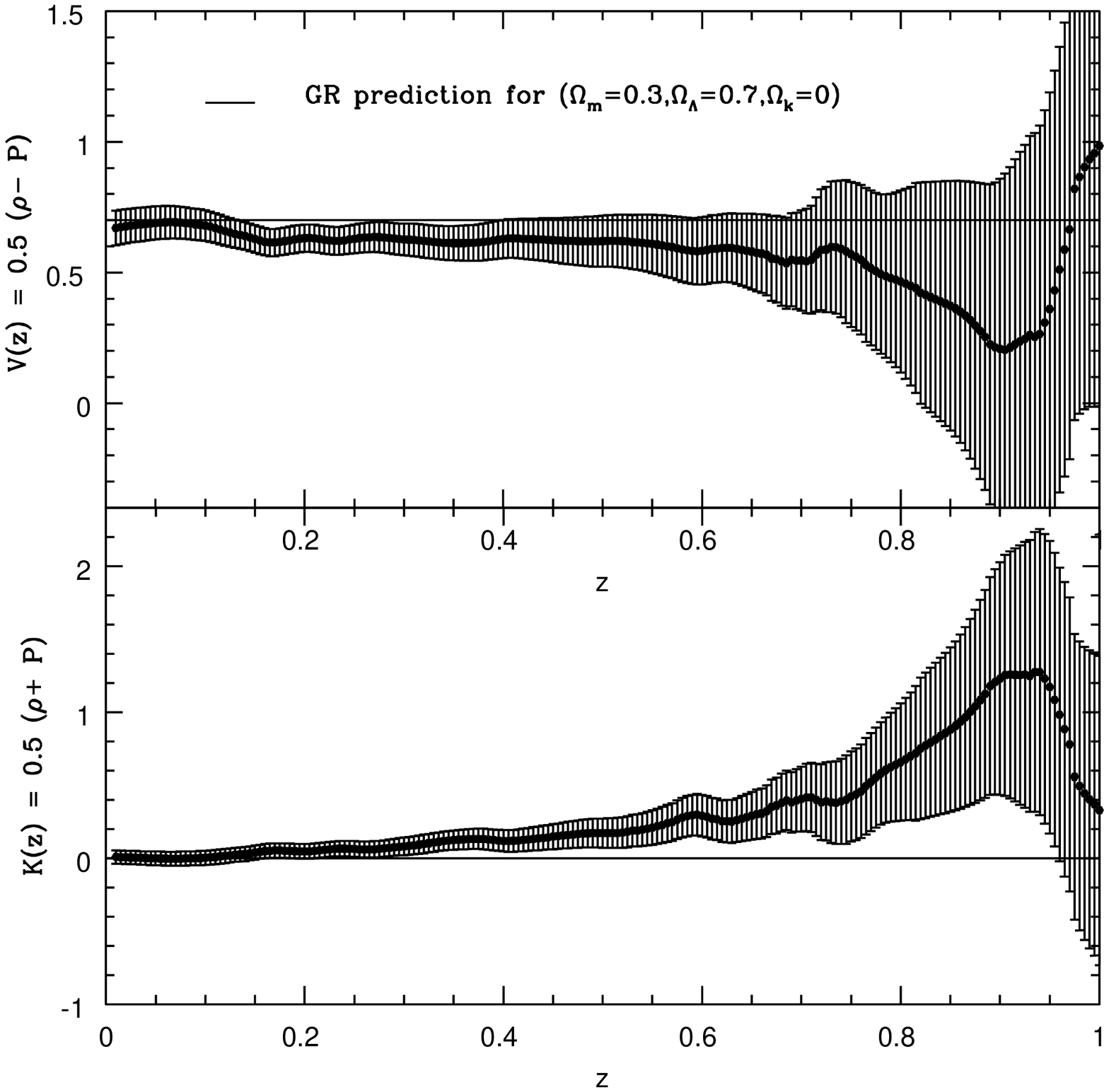}
\label{KV}
  \caption{The kinetic and potential energy densities of
a dark energy scalar field can be obtained by combining
the pressure and energy density of the dark energy.  These
are shown here for the supernovae sample of \cite{R07}.}
\end{figure}

To study the properties of the dark energy as a function
of redshift, the first and second derivatives
of the coordinate distance can be combined 
to solve for the pressure, energy density, equation of
state, and kinetic and potential energy densities of the 
dark energy if a theory
of gravity be specified, as discussed by 
\cite{DD04}. Einstein's Equations from General Relativity 
for an isotropic and homogeneous universe 
are used to define the relationship between
the dark energy pressure and energy density, 
the mass-energy density of non-relativistic matter, and the 
expansion and acceleration rates of the universe.  
Given that the expansion and acceleration rates of
the universe are known independent of the contents of
the universe, these equations allow a determination of
the pressure, energy density, and equation of state of the
dark energy as functions of redshift, as shown by \cite{DD04},
and the kinetic and potential energy density of the dark energy.  

\section{Summary}

Our model-independent method 
has been applied to the 182 SN presented by \cite{R07}.
The values of the coordinate distance to each source are obtained
from the distance moduli, as described by \cite{D07A}.  

As shown in Figure \ref{yeq}, the current value of the
acceleration parameter is 
$q_0 = -0.5 \pm 0.13$.  This indicates that the universe is accelerating 
today with greater than $3 \sigma$ confidence.  
Note that our
determination of $q_0$ depends only upon the assumption that 
the universe is described by a Robertson-Walker metric, and
{\it is independent of the value of space curvature}, as discussed 
in detail by \cite{D07B}, and as is clear from the equations
presented by \cite{DD03}. 

The redshift of transition from a state of acceleration to 
a state of deceleration does require that a value of space
curvature be specified, so two assumption are made to 
determine $z_T$: zero space curvature and a Robertson-Walker 
line element. For the sample of \cite{R07}, we find
$z_T = 0.35 {}^{+0.15}_{-0.7}$.  

The first and second derivatives of the
coordinate distance with respect to redshift 
are combined 
to solve for
the pressure, energy density, and equation of state of
the dark energy as functions of redshift by applying 
the equations derived by \cite{DD04}. The results are
shown in Figure \ref{Pfw}. 
The first and second derivatives of the coordinate distance
can also be combined to obtain the kinetic and potential
energy density of a dark energy scalar field, and 
these results are shown in Figure \ref{KV}.  

Overall, the results are consistent with those expected 
in a standard Lambda Cold Dark Matter model, in which 
a cosmological constant is responsible for the acceleration 
history of the universe, at about the two sigma level. 
Further details of this study
are presented by \cite{D07B}.



\begin{theacknowledgments}
We would like to thank the observers for their tireless efforts
to obtain the supernovae data that makes these studies possible.
This work was supported in part by U. S. National Science
Foundation grants AST-0507465 (R.A.D.) and AST-0407448 (S.G.D.),
and the Ajax Foundation (S.G.D.).   
\end{theacknowledgments}







\end{document}